\documentclass[12pt]{article}
\usepackage{amsmath}
\usepackage{amssymb}

\newcommand{\chh}{{\cal H}}
\newcommand{\field}[1]{\mathbb{#1}}
\newcommand{\R}{\field{R}}
\newcommand{\crr}{{\cal R}}
\newcommand{\cii}{{\cal I}}
\newcommand{\bp}{b^{\dagger}}

\title{
COMMENTS ON DIHEDRAL AND SUPERSYMMETRIC EXTENSIONS OF A FAMILY OF HAMILTONIANS ON A PLANE}
\author{C. QUESNE\\
{\small \sl Physique Nucl\'eaire Th\'eorique et Physique Math\'ematique,}
\\ {\small \sl Universit\'e Libre de Bruxelles, Campus de la Plaine CP229,} \\ 
{\small \sl Boulevard~du Triomphe, B-1050 Brussels, Belgium} \\
{\small \sl cquesne@ulb.ac.be}}
\date{ }
\begin{document}
\baselineskip=22pt plus 1pt minus 1pt
\maketitle

\begin{abstract} 
{}For any odd $k$, a connection is established between the dihedral and supersymmetric extensions of the Tremblay-Turbiner-Winternitz Hamiltonians $H_k$ on a plane. For this purpose, the elements of the dihedral group $D_{2k}$ are realized in terms of two independent pairs of fermionic creation and annihilation operators and some interesting trigonometric identities are demonstrated.    
\end{abstract}

\noindent
Running head: Dihedral and Supersymmetric Extensions


%
%
\newpage
\section{Introduction}

Recently, Tremblay, Turbiner and Winternitz (TTW) introduced an infinite family of exactly solvable and integrable Hamiltonians on a plane \cite{tremblay09}
\begin{equation}
\begin{split}
  & H_k = - \partial_r^2 - \frac{1}{r} \partial_r - \frac{1}{r^2} \partial_{\varphi}^2 + \omega^2 r^2 +
       \frac{k^2}{r^2} [a(a-1) \sec^2 k\varphi + b(b-1) \csc^2 k\varphi], \\
  & 0 \le r < \infty, \qquad 0 \le \varphi < \frac{\pi}{2k},
\end{split}  \label{eq:TTW}
\end{equation}
including as specials cases several well-known Hamiltonians, such as those of the Smorodinsky-Winternitz system ($k=1$) \cite{fris, winternitz}, of the rational $BC_2$ model ($k=2$) \cite{olsha}, and of the three-particle Calogero model \cite{calogero69} with some extra three-body interaction ($k=3$) \cite{wolfes, calogero74}.\par
%
%
This new family has aroused a lot of interest due to the conjecture made in \cite{tremblay09} that $H_k$ is superintegrable for any integer value of $k$. Several works have therefore been devoted to proving such a conjecture first for the classical Hamiltonians \cite{tremblay10, kalnins09, kalnins10a}, then for the quantum ones \cite{cq10a, kalnins10b}. In the demonstration performed in \cite{cq10a} and valid for any odd $k$, use has been made of a realization of the dihedral groups $D_{2k}$ on the plane, leading to $D_{2k}$-extended and invariant Hamiltonians $\chh_k$ \cite{cq10b}. For the latter, a Dunkl operator formalism previously employed for $k=3$ \cite{cq95} has proved to provide a very convenient tool.\par
%
%
More recently, another type of extension of $H_k$ has been considered in terms of two independent pairs of fermionic creation and annihilation operators, yielding supersymmetrized Hamiltonians $\chh^s$ as well as corresponding supercharges \cite{cq10c}. Such an approach gives rise to a dynamical ${\rm osp}(2/2, \R) \sim {\rm su}(1,1/1)$ superalgebra of the same kind as that introduced by Freedman and Mende for the Calogero problem \cite{freedman} and studied later on by several authors (see, e.g., \cite{brink93, brink98, ghosh01, ghosh04, gala}).\par
%
%
The purpose of the present work is to comment on the relations existing between both extensions of $H_k$.\par
%
%
\section{\boldmath Dihedral versus Supersymmetric Extensions of $H_k$}

The dihedral group $D_{2k}$, where $k$ may be any positive integer, is known to have $4k$ elements $\crr^i$ and $\crr^i \cii$, $i=0$, 1, \ldots,~$2k-1$, satisfying the relations
\begin{equation}
  \crr^{2k} = 1, \qquad \cii^2 = 1, \qquad \cii \crr = \crr^{2k-1} \cii,  \label{eq:def-1}
\end{equation}
and, in the case of a unitary representation, the Hermiticity conditions
\begin{equation}
  \crr^{\dagger} = \crr^{2k-1}, \qquad \cii^{\dagger} = \cii.  \label{eq:def-2}
\end{equation}
In the context of the TTW Hamiltonians, $\crr$ and $\cii$ are realized by some operators acting in the plane $(r, \varphi)$, namely $\crr = \exp\left(\frac{1}{k} \pi \partial_{\varphi}\right)$ is the rotation operator through angle $\pi/k$ while $\cii = \exp({\rm i} \pi \varphi \partial_{\varphi})$ changes $\varphi$ into $-\varphi$.\par
%
%
In \cite{cq10b}, it has been shown that for any odd $k$, $H_k$ can be extended by incorporating some of the $D_{2k}$ elements in such a way that the resulting Hamiltonian
\begin{equation}
\begin{split}
  \chh_k & = - \partial_r^2 - \frac{1}{r} \partial_r - \frac{1}{r^2} \partial_{\varphi}^2 + \omega^2 r^2 +
      \frac{1}{r^2} \biggl[\sum_{i=0}^{k-1} \sec^2 \left(\varphi + \frac{i\pi}{k}\right) a (a - \crr^{k+2i} \cii) \\
  & \quad + \sum_{i=0}^{k-1} \csc^2 \left(\varphi + \frac{i\pi}{k}\right) b (b - \crr^{2i} \cii)\biggr]
\end{split}  \label{eq:ext-H}
\end{equation}
remains invariant under $D_{2k}$, while giving back $H_k$ by projection in the representation space of the $D_{2k}$ identity representation, i.e., by replacing $\crr$ and $\cii$ by 1. The last point is due to the existence of the two trigonometric identities \cite{jakubsky}
\begin{equation}
  \sum_{i=0}^{k-1} \sec^2 \left(\varphi + \frac{i \pi}{k}\right) = k^2 \sec^2 k\varphi, \qquad 
  \sum_{i=0}^{k-1} \csc^2 \left(\varphi + \frac{i \pi}{k}\right) = k^2 \csc^2 k\varphi,  \label{eq:identities}
\end{equation}
valid for any odd $k$, from which it follows that
\begin{equation}
  \chh_k - H_k = \frac{1}{r^2} \biggl[a \sum_{i=0}^{k-1} \sec^2 \left(\varphi + \frac{i\pi}{k}\right) (1 
  - \crr^{k+2i} \cii) + b \sum_{i=0}^{k-1} \csc^2 \left(\varphi + \frac{i\pi}{k}\right) (1 - \crr^{2i} \cii)\biggr].  
  \label{eq:difference}
\end{equation}
\par
%
%
{}For even $k$ values, there also exist some $D_{2k}$-extended Hamiltonians $\chh_k$ with similar properties \cite{cq10b}. However, the trick employed to compensate for lack of suitable counterpart of the first identity in (\ref{eq:identities}) turns out to be rather artificial so that up to now such $\chh_k$'s have not proved useful to demonstrate the superintegrability of $H_k$ for even $k$. We shall therefore not consider them here.\par
%
%
On the other hand, on introducing two pairs of fermionic creation and annihilation operators $(\bp_x, b_x)$ and $(\bp_y, b_y)$, such that $\{b_x, \bp_x\} = \{b_y, \bp_y\} = 1$ while the remaining anticommutators vanish, $H_k$ can be extended to some supersymmetrized Hamiltonian
\begin{equation}
  {\cal H}^s = H_{k,{\rm B}} + H_{k,{\rm F}}, \qquad H_{k,{\rm B}} = H_k, \qquad H_{k,{\rm F}} = 4\omega
  (\Gamma + Y),  \label{eq:super-H}
\end{equation}
for any positive real value of $k$ \cite{cq10c}. In (\ref{eq:super-H}), the operators $4 \omega \Gamma$ and $4 \omega Y$ are given by
\begin{equation}
\begin{split}
  4 \omega \Gamma & = \frac{2k}{r^2}\biggl\{a \sec^2 k\varphi \biggl[\Bigl(\cos (k-2)\varphi 
       \cos k\varphi + \frac{k}{2}(1 - \cos 2\varphi)\Bigr) \bp_x b_x \\  
  & \quad - \Bigl(\sin (k-2)\varphi \cos k\varphi + \frac{k}{2} \sin 2\varphi\Bigr) (\bp_x b_y + \bp_y b_x) \\
  & \quad + \Bigl(- \cos (k-2)\varphi \cos k\varphi + \frac{k}{2}(1 + \cos 2\varphi)\Bigr) \bp_y b_y\biggr] \\ 
  & \quad + b \csc^2 k\varphi \biggl[\Bigl(\sin (k-2) \varphi \sin k\varphi + \frac{k}{2}(1 - \cos 2\varphi)\Bigr)
       \bp_x b_x \\ 
  & \quad + \Bigl(\cos (k-2)\varphi \sin k\varphi - \frac{k}{2} \sin 2\varphi\Bigr) (\bp_x b_y + \bp_y b_x) \\ 
  & \quad + \Bigl(- \sin (k-2)\varphi \sin k\varphi + \frac{k}{2}(1 + \cos 2\varphi)\Bigr) \bp_y b_y\biggr]
      \biggr\} 
\end{split}  \label{eq:gamma}
\end{equation}
and
\begin{equation}
  4 \omega Y = 2 \omega [\bp_x b_x + \bp_y b_y - k(a+b) - 1],  
\end{equation}
respectively.\par
%
%
In the remainder of this paper, we plan to show that although at first sight the operators $\chh_k - H_k$ and $4 \omega \Gamma$, defined in (\ref{eq:difference}) and (\ref{eq:gamma}), look very different, one can establish some relationship between them.\par
%
%
\section{\boldmath Realization of the $D_{2k}$ Elements in Terms of Fermionic Operators}

As a first step, let us prove that for any integer $k$, the $D_{2k}$ elements $\crr^i$ and $\crr^i \cii$, $i=0$, 1, \ldots,~$2k-1$, can be realized in terms of two independent pairs of fermionic operators $(\bp_x, b_x)$ and $(\bp_y, b_y)$.\par
%
%
{}For such a purpose, let us define $\crr$ and $\cii$ as
\begin{equation}
  \crr \equiv 1 + \left(\cos \frac{\pi}{k} - 1\right) (\bp_x b_x + \bp_y b_y) + \sin \frac{\pi}{k} (\bp_x b_y
  - \bp_y b_x) + 2 \left(1 - \cos \frac{\pi}{k}\right) \bp_x b_x \bp_y b_y  \label{eq:R} 
\end{equation}
and
\begin{equation}
  \cii \equiv 1 - 2 \bp_y b_y = - [\bp_y, b_y],  \label{eq:I}
\end{equation}
respectively. As it is obvious, the latter satisfies the second relations in (\ref{eq:def-1}) and (\ref{eq:def-2}).\par
%
%
On starting from $\crr$ given in (\ref{eq:R}), it is then easy to prove by induction over $i$ that
\begin{equation}
  \crr^i = 1 + \left(\cos \frac{i\pi}{k} - 1\right) (\bp_x b_x + \bp_y b_y) + \sin \frac{i\pi}{k} (\bp_x b_y
  - \bp_y b_x) + 2 \left(1 - \cos \frac{i\pi}{k}\right) \bp_x b_x \bp_y b_y  \label{eq:R-i} 
\end{equation}
for any $i=0$, 1, \ldots,~$2k-1$ and that $\crr^{2k} = 1$ in agreement with the first relation in (\ref{eq:def-1}). Moreover
\begin{equation*}
  \crr^{2k-1} = 1 + \left(\cos \frac{\pi}{k} - 1\right) (\bp_x b_x + \bp_y b_y) - \sin \frac{\pi}{k} (\bp_x 
  b_y - \bp_y b_x) + 2 \left(1 - \cos \frac{\pi}{k}\right) \bp_x b_x \bp_y b_y 
\end{equation*}
coincides with $\crr^{\dagger}$, thus proving the first relation in (\ref{eq:def-2}).\par
%
%
Next, on multiplying (\ref{eq:R-i}) by (\ref{eq:I}), we get
\begin{equation}
  \crr^i \cii = 1 + \left(\cos \frac{i\pi}{k} - 1\right) \bp_x b_x - \left(\cos \frac{i\pi}{k} + 1\right) \bp_y b_y
  - \sin \frac{i\pi}{k} (\bp_x b_y + \bp_y b_x)  \label{eq:R-i-I} 
\end{equation}
for $i=0$, 1, \ldots,~$2k-1$. We may note in passing that $\crr^k \cii$ reduces to
\begin{equation}
  \crr^k \cii = 1 - 2 \bp_x b_x = - [\bp_x, b_x].  \label{eq:R-k-I}
\end{equation}
Furthermore, it results from (\ref{eq:R-i-I}) that
\begin{equation*}
  \crr^{2k-1} \cii = 1 + \left(\cos \frac{\pi}{k} - 1\right) \bp_x b_x - \left(\cos \frac{\pi}{k} + 1\right) \bp_y 
  b_y + \sin \frac{\pi}{k} (\bp_x b_y + \bp_y b_x), 
\end{equation*}
which can be easily shown to be identical with $\cii \crr$ obtained from (\ref{eq:R}) and (\ref{eq:I}). Hence, the third relation in (\ref{eq:def-1}) is also demonstrated, which completes the proof.\par
%
%
It is worth observing that the elements $\crr^i \cii$, $i=0$, 1, \ldots,~$2k-1$, which are the only ones appearing in (\ref{eq:ext-H}) and (\ref{eq:difference}), can be rewritten in a very compact way as
\begin{equation}
  \crr^i \cii = 1 - 2 \bp_i b_i = - [\bp_i, b_i]  \label{eq:R-i-I-bis}
\end{equation}
in terms of some `rotated' creation and annihilation operators
\begin{equation}
  \bp_i \equiv \sin \frac{i\pi}{2k} \bp_x + \cos \frac{i\pi}{2k} \bp_y, \qquad b_i \equiv \sin \frac{i\pi}{2k} 
  b_x + \cos \frac{i\pi}{2k} b_y, \qquad i=0, 1, \ldots, 2k-1,  \label{eq:b-i} 
\end{equation}
satisfying the deformed anticommutation relations
\begin{equation*}
  \{b_i, b_j\} = \{\bp_i, \bp_j\} = 0, \qquad \{b_i, \bp_j\} = \cos \frac{(j-i)\pi}{2k}.
\end{equation*}
\par
%
%
As a final point, let us mention that the realization given in Eqs.\ (\ref{eq:R}) and (\ref{eq:I}) is only valid at the operator level, but not as far as the action of operators on wavefunctions is concerned: the operators $\crr$ and $\cii$, as considered in \cite{cq10b}, were acting on functions $\psi(r, \varphi)$ defined in the plane, whereas the operators on the right-hand side of (\ref{eq:R}) and (\ref{eq:I}) act on the fermionic degrees of freedom used to extend the configuration space in the supersymmetric approach.\par
%
%
\section{\boldmath Comparison between Both Extensions of $H_k$}

{}For odd $k$, let us now insert the realization (\ref{eq:R-i-I-bis}) of $\crr^i \cii$ in terms of fermionic operators in Eq.\ (\ref{eq:difference}). The resulting operator reads
\begin{equation*}
  4 \omega \tilde{\Gamma} = \frac{2}{r^2} \biggl[a \sum_{i=0}^{k-1} \sec^2 \left(\varphi + \frac{i\pi}{k}\right) 
  \bp_{k+2i} b_{k+2i} + b \sum_{i=0}^{k-1} \csc^2 \left(\varphi + \frac{i\pi}{k}\right) \bp_{2i} b_{2i}\biggr]  
\end{equation*}
or, by taking definition (\ref{eq:b-i}) into account,
\begin{equation*}
\begin{split}
  4 \omega \tilde{\Gamma} & = \frac{a}{r^2} \sum_{i=0}^{k-1} \sec^2 \left(\varphi + \frac{i\pi}{k}\right) 
     \biggl[\biggl(\cos \frac{2i\pi}{k} + 1\biggr) \bp_x b_x - \sin \frac{2i\pi}{k} (\bp_x b_y + \bp_y b_x) \\
  & \quad + \biggl(- \cos \frac{2i\pi}{k} + 1\biggr) \bp_y b_y\biggr] \\
  & \quad + \frac{b}{r^2} \sum_{i=0}^{k-1} \csc^2 \left(\varphi + \frac{i\pi}{k}\right) \biggl[\biggl(- \cos 
      \frac{2i\pi}{k} + 1\biggr) \bp_x b_x + \sin \frac{2i\pi}{k} (\bp_x b_y + \bp_y b_x) \\
  & \quad + \biggl(\cos \frac{2i\pi}{k} + 1\biggr) \bp_y b_y \biggr].   
\end{split}
\end{equation*}
\par
%
%
Such an operator will coincide with the operator $4 \omega \Gamma$, coming from the supersymmetric extension and given in (\ref{eq:gamma}), provided the coefficients of $a \bp_x b_x$, $a (\bp_x b_y + \bp_y b_x)$, $a \bp_y b_y$, $b \bp_x b_x$, $b (\bp_x b_y + \bp_y b_x)$, and $b \bp_y b_y$ are equal. Among the resulting six equations, there only remain two independent ones when use is made of the identities (\ref{eq:identities}) and of some simple properties of trigonometric functions. They may be written as
\begin{equation}
\begin{split}
  & \sum_{i=0}^{k-1} \sec^2 \left(\varphi + \frac{i\pi}{k}\right) \cos \frac{2i\pi}{k} = k \sec^2 k\varphi
        \{\cos[2(k-1)\varphi] - (k-1) \cos 2\varphi\}, \\
  & \sum_{i=0}^{k-1} \sec^2 \left(\varphi + \frac{i\pi}{k}\right) \sin \frac{2i\pi}{k} = k \sec^2 k\varphi
        \{\sin[2(k-1)\varphi] + (k-1) \sin 2\varphi\}.
\end{split}  \label{eq:identities-bis}
\end{equation}
We plan to show that these relations are trigonometric identities.\par
%
%
{}For such a purpose, it proves useful to first integrate them to get some simpler relations. The results read
\begin{equation}
\begin{split}
  & \sum_{i=0}^{k-1} \tan \left(\varphi + \frac{i\pi}{k}\right) \cos \frac{2i\pi}{k} = - k \frac{\sin[(k-2)\varphi]}
       {\cos k\varphi}, \\
  & \sum_{i=0}^{k-1} \tan \left(\varphi + \frac{i\pi}{k}\right) \sin \frac{2i\pi}{k} = k \frac{\cos[(k-2)\varphi]}
       {\cos k\varphi} - \delta_{k,1},
\end{split}  \label{eq:identities-ter}
\end{equation}
where the integration constants have been determined from the values of both sides for $\varphi = 0$. It is then convenient to express $\cos \frac{2i\pi}{k}$ and $\sin \frac{2i\pi}{k}$ on the left-hand sides of these equations as
\begin{equation*}
\begin{split}
  & \cos \frac{2i\pi}{k} = \cos \left(2\varphi + \frac{2i\pi}{k}\right) \cos 2\varphi + \sin \left(2\varphi + 
        \frac{2i\pi}{k}\right) \sin 2\varphi, \\
  & \sin \frac{2i\pi}{k} = \sin \left(2\varphi + \frac{2i\pi}{k}\right) \cos 2\varphi - \cos \left(2\varphi + 
        \frac{2i\pi}{k}\right) \sin 2\varphi. 
\end{split}
\end{equation*}
On taking into account that
\begin{equation*}
\begin{split}
  & \sum_{i=0}^{k-1} \tan \left(\varphi + \frac{i\pi}{k}\right) \sin \left(2\varphi + \frac{2i\pi}{k}\right) =
        2 \sum_{i=0}^{k-1} \sin^2 \left(\varphi + \frac{i\pi}{k}\right) \\
  & \quad = \sum_{i=0}^{k-1} \left[1 - \cos \left(2\varphi + \frac{2i\pi}{k}\right)\right] = k - \delta_{k,1}
        \cos 2\varphi 
\end{split}
\end{equation*}
and
\begin{equation*}
\begin{split}
  & \sum_{i=0}^{k-1} \tan \left(\varphi + \frac{i\pi}{k}\right) \cos \left(2\varphi + \frac{2i\pi}{k}\right)
          = \sum_{i=0}^{k-1} \tan \left(\varphi + \frac{i\pi}{k}\right) \left[2 \cos^2 \left(\varphi + 
          \frac{i\pi}{k}\right) - 1 \right] \\
  & \quad = \sum_{i=0}^{k-1} \left[\sin \left(2\varphi + \frac{2i\pi}{k}\right) - \tan \left(\varphi + \frac{i\pi}{k}
          \right)\right] = \delta_{k,1} \sin 2\varphi - k \tan k\varphi,
\end{split}
\end{equation*}
where in the last step use has been made of Eq.\ (4.4.7.1) of \cite{prudnikov}, it is straightforward to show that Eq.\ (\ref{eq:identities-ter}) is indeed satisfied, which completes the proof of (\ref{eq:identities-bis}).\par
%
%
As a final point, it is worth observing that the complete supersymmetric Hamiltonian $\chh^s$, given in (\ref{eq:super-H}), would be obtained by adding the operator $- 2 \omega \left[\frac{1}{2} (1 + \crr^k) \cii + k(a +
b)\right]$ to the $D_{2k}$-extended Hamiltonian $\chh_k$ and employing Eqs.\ (\ref{eq:I}) and (\ref{eq:R-k-I}).\par
%
%
\section{Conclusion}

Here we have demonstrated that for any odd $k$ there exists some connection between the $D_{2k}$ and supersymmetric extensions of the TTW Hamiltonians on a plane, introduced in \cite{cq10b} and \cite{cq10c}, respectively. Such a relation has been obtained by realizing the $D_{2k}$ elements in terms of two independent pairs of fermionic creation and annihilation operators and by establishing some trigonometric identities, which, as far as the author knows, are new and might be of some interest in other contexts.\par
%
%
The relationship observed between the two extensions of the TTW Hamiltonians may be considered as a generalization of another one previously noted \cite{wyllard} between the $S_n$ extension of the $n$-particle Calogero model \cite{poly, brink92} and the supersymmetric one of the latter. As such, it hints at a possible existence of similar connections valid for all Calogero- and Sutherland-type models associated with Lie algebra root systems. This might provide an interesting issue for future investigation.\par
%
%
\section*{Acknowledgments}

The author would like to thank M.\ S.\ Plyushchay for some interesting comments on connections with some of his works, e.g.\ the one published in {\em Ann.\ Phys.\ (N.\ Y.)} {\bf 245}, 339 (1996).\par
%
%

\end{document}